\RequirePackage[T1]{fontenc}
\documentclass[10pt,conference]{IEEEtran}

\usepackage{booktabs}
\usepackage{array}
\usepackage{xurl}
\usepackage[hidelinks]{hyperref}

\title{Yesterday's Interviews for Today's Engineers:\\
Retrospective Perceptions and a Work-Aligned Hiring Framework (2005--2026)}

\author{\IEEEauthorblockN{Vitalii Romaniuk}
\IEEEauthorblockA{Independent Researcher\\
Email: kidomakai@gmail.com}}

\begin{document}

\maketitle

\begin{abstract}
Technical hiring has increasingly become a skill system of its own, requiring candidates to prepare for processes that only partially resemble the position being filled. This paper reconstructs how the software-engineering community remembers that process across four eras from 2005 to 2026. Retrospective ratings from 911 respondents show a history of expanding interview work and burden, while an analysis of public preparation questions finds that most tested knowledge has a plausible use in an ordinary software-engineering project. The contradiction is only apparent: relevant topics can still form an invalid process when they are accumulated beyond the vacancy, presented under artificial constraints, or selected and interpreted without a common job-derived standard. The paper responds with the Ticket-Based Pragmatic Assessment Framework, a prescriptive reference framework that derives a bounded assessment from declared responsibilities, near-term tickets, and project-entry work. It preserves normal tools, defines three explicit evidence gates, limits repeated sessions, and stops when the decision is supported. The objective is to make an interview a compressed version of the job rather than a parallel profession that candidates must learn merely to obtain it.
\end{abstract}


\section{Introduction}
\label{sec:introduction}

Technical hiring affects both organizational capability and candidate access to software engineering careers. Organizations invest substantial resources in identifying capable engineers, while candidates must navigate processes that may include online assessments, live coding, take-home assignments, system-design interviews, and behavioral evaluations.

During the same period, however, software engineering itself has undergone profound transformation. Daily engineering work now routinely involves distributed systems, cloud services, legacy code maintenance, collaborative development, automated deployment pipelines, and AI-assisted programming. These changes raise an important empirical question: have technical interview practices evolved alongside software engineering practice, or have the two followed different trajectories?

Although prior studies have examined interview anxiety, candidate experience, and assessment validity, relatively little work has investigated how the software-engineering community remembers changes in interviews over an extended historical period. That remembered history matters in its own right: hiring practices are experienced, discussed, taught to later candidates, and compared with earlier professional norms. A retrospective comparison cannot recreate a contemporaneous archive, but it can identify the historical pattern that remains in collective professional judgment.

To address this gap, this paper reconstructs a remembered history of interview practice across four eras spanning 2005 to 2026. The empirical study combines retrospective ratings from 911 respondents recruited through software-engineering communities with an audited corpus of public interview-preparation prompts. The survey captures the comparative historical pattern of process structure, burden, and relevance retained by respondents; the benchmark examines whether the knowledge tested by public preparation material has a plausible use in an ordinary software-engineering project. The paper then converts the combined findings into a concrete reference design for a shorter, role-derived assessment. Section~\ref{sec:research_questions} states the four research questions.

This paper makes the following contributions:

\begin{enumerate}
\item A retrospective empirical reconstruction of the professional history remembered across four software-engineering interview eras.

\item A repeated-measures survey of 911 software-engineering community respondents covering interview duration, number of rounds, perceived burden, and interview--work disconnect.

\item An audited 945-prompt benchmark that maps public interview-preparation questions to activities performed in modern software engineering.

\item The Ticket-Based Pragmatic Assessment Framework, which translates role responsibilities, near-term tickets, and project-entry work into a bounded, tool-authentic assessment and an explicit evaluation agenda.
\end{enumerate}

\section{Related Work}
\label{sec:related_work}

Technical hiring has attracted increasing attention within empirical software engineering over the last decade. Existing research has examined technical interviews from several complementary perspectives, including candidate experience, interview performance, hiring processes, assessment validity, and personnel selection theory. Collectively, these studies provide valuable insight into individual aspects of technical hiring; however, they generally investigate isolated dimensions of the recruitment process rather than its long-term evolution. This section reviews the most relevant literature and positions the contribution of the present study.

\subsection{Research on Technical Interviews}

Software engineering interviews have been studied primarily from the perspective of candidate experience and hiring effectiveness. Previous work has explored how developers perceive technical interviews, identifying recurring concerns regarding preparation effort, realism of interview tasks, evaluation transparency, consistency of interviewer expectations, and fairness throughout the hiring process \cite{behroozi2019hiring,ford2017techtalk,behroozi2020debugging}. These studies suggest that technical interviews represent considerably more than a programming assessment; rather, they combine technical problem solving with communication, reasoning, and interpersonal interaction.

Large-scale analyses of public hiring reviews further demonstrate that candidate experience is influenced not only by technical assessment itself but also by communication quality, interview organization, interviewer consistency, and transparency of evaluation procedures \cite{behroozi2020debugging}. Collectively, these findings have motivated continued research into improving technical hiring while maintaining reliable candidate assessment.

Preparation research provides a complementary view of this burden. In a survey of 131 candidates actively preparing for software engineering interviews, Bell et al. found that candidates used many resources but rarely practiced in authentic settings that combined coding, communication, and observation; participants consequently reported anxiety and limited preparedness \cite{bell2025prepare}. A qualitative study of 16 minoritized computing students similarly identified uncertainty, preparation time, anxiety management, interview technique, and inclusivity as connected obstacles in the path to employment \cite{lunn2022need}. These studies indicate that candidate burden begins before the interview itself and may be distributed unevenly according to access to preparation support.

Applicant reactions are also consequential for organizations. A meta-analysis of 86 samples involving 48,750 participants found that favorable perceptions of selection procedures were associated with organizational attractiveness, offer-acceptance intentions, and willingness to recommend the employer. Interviews and work samples were generally viewed more favorably than less job-related assessment methods \cite{hausknecht2004reactions}. This evidence provides a counterpoint to treating candidate experience and assessment quality as competing objectives: job-related procedures may support both.

\subsection{Human Factors During Technical Interviews}

Another important research direction concerns the human factors influencing interview performance. Experimental studies have shown that observation during live coding interviews can substantially increase stress while simultaneously reducing technical performance \cite{behroozi2020does}. In an earlier 11-participant study, Behroozi et al. reported preliminary evidence of higher cognitive-load indicators during whiteboard problem solving than during an equivalent paper condition, suggesting that interview format can influence performance beyond technical competence alone \cite{behroozi2018dazed}. These findings are consistent with broader cognitive load theory, which explains how excessive mental workload can negatively affect reasoning and problem solving \cite{sweller1988cognitive}.

More recent research has investigated alternative interview formats designed to reduce unnecessary cognitive load while preserving assessment quality. Asynchronous interview approaches have shown promise in reducing candidate anxiety while maintaining opportunities to evaluate technical reasoning and communication skills \cite{behroozi2022async}. Together, these studies demonstrate that interview design itself substantially influences candidate performance and should therefore be considered when evaluating hiring methodologies.

\subsection{Assessment Validity}

Outside software engineering, personnel psychology has extensively investigated employee selection methods. A widely cited meta-analysis placed structured interviews and work-sample assessments among the strongest predictors of future job performance \cite{schmidt1998validity}. More recent work, however, identified systematic overcorrection for range restriction in earlier meta-analytic estimates and produced substantially revised validity values \cite{sackett2022revisiting}. The updated evidence still ranks structured interviews highly, but cautions against treating older effect-size estimates or simple rankings as settled facts.

The literature is clearer about process design. Reviews of structured interviewing show that standardization can improve reliability and comparability across candidates \cite{campion1997structure,levashina2014structured}. Highhouse and Brooks further argue that selection noise can be reduced by decomposing judgments, agreeing on standards in advance, and aggregating independent evaluations \cite{highhouse2023noise}. Accordingly, the relevant contrast is not simply ``interviews versus no interviews,'' but between assessments that are job-related and consistently scored and those that rely on loosely defined or interviewer-specific judgments.

Although these principles are well established within personnel selection research, comparatively few studies have examined how they have been adopted within software engineering interviews or how interview practices have evolved alongside changes in software engineering itself.

\subsection{Evolution of Software Engineering Practice}

Software engineering has undergone substantial transformation since the early 2000s. Cloud computing, distributed services, DevOps, and continuous-delivery practices became prominent parts of modern software development and operation \cite{armbrust2010view,kim2016devops}. More recently, large language models and AI-assisted development tools have begun reshaping software engineering workflows, introducing new patterns of software creation, maintenance, and developer productivity \cite{hou2024large}.

These technological advances have fundamentally changed the knowledge, tools, and workflows expected of professional software engineers. Consequently, evaluating whether technical interviews continue to assess competencies relevant to contemporary engineering practice represents an important empirical question.

\subsection{Research Gap}

Existing research has substantially advanced understanding of software engineering hiring by investigating candidate perceptions, interview stress, cognitive load, hiring pipelines, interviewer expectations, and assessment validity. Nevertheless, several important gaps remain.

First, previous studies have primarily investigated contemporary interview practices rather than retrospective perceptions across multiple technological eras. Second, most work has focused on either candidate experiences or organizational hiring practices, whereas comparatively little research combines the perspectives of software engineers and hiring managers within one historical comparison. Third, although concerns regarding interview realism have frequently been discussed, relatively little empirical work has systematically analyzed the competencies emphasized by public interview-preparation questions and compared them with competencies used in contemporary software engineering practice.

To address these gaps, the present study examines retrospective perceptions of software engineering interview practices across four eras spanning 2005 to 2026. It combines ratings from 911 software-engineering community respondents with an audited public preparation corpus, providing two perspectives on the relationship between technical hiring and modern software engineering practice.

\section{Research Questions}
\label{sec:research_questions}

This study investigates the history of software engineering interviews as reconstructed in present-day professional memory and asks whether that history reflects the evolution of contemporary software engineering practice. The four eras are 2005--2010, 2011--2017, 2018--2022, and 2023--2026. Building upon the literature reviewed in the previous section, the study addresses four research questions.

\subsection*{RQ1: Procedural Expansion}

\textit{How do respondents perceive live-interview duration and the number of interview rounds across the four historical development eras (2005--2010, 2011--2017, 2018--2022, and 2023--2026)?}

This question isolates the observable structure of the process: how much live time it consumes and how many separately scheduled stages it contains.

\vspace{0.5em}

\subsection*{RQ2: Disconnect and Burden}

\textit{How do perceived interview--work disconnect and combined psychological and time burden differ across the four eras?}

This question examines whether greater procedure is accompanied by a greater experiential cost and whether perceived job relevance follows the same trajectory.

\vspace{0.5em}

\subsection*{RQ3: Developer and Hiring Manager Perspectives}

\textit{How do respondents identifying as software engineers or candidates and those identifying as engineering managers compare in their ratings of contemporary technical interviews?}

This question provides a descriptive comparison of Era~IV duration, rounds, disconnect, and burden. It does not independently measure hiring effectiveness or assessment quality.

\vspace{0.5em}

\subsection*{RQ4: Interview Content and Professional Practice}

\textit{To what extent do prompts in a public interview-preparation corpus test knowledge with a plausible application in an ordinary contemporary software-engineering project?}

This research question complements the survey by examining the content candidates encounter in public preparation material. Its reference point is broad professional practice: whether the tested knowledge could reasonably arise in a typical software project, not whether the exact prompt would be spoken during daily work. It does not estimate how often a prompt is used by employers, whether it predicts job performance, or whether it is appropriate for a particular vacancy, role, or seniority level.

\section{Methodology}
\label{sec:methodology}

This study used a retrospective, repeated-measures survey supplemented by a benchmark analysis of public interview-preparation prompts. The objective was to reconstruct the professional history remembered for software-engineering interviews between 2005 and 2026 and to compare preparation content with contemporary engineering practice.

\subsection{Research Design}

The study was designed to address the four research questions in Section~\ref{sec:research_questions}. Each respondent evaluated all four historical eras, producing paired observations for within-respondent comparison. This retrospective structure was intentional. Asking participants to rate every era within a common present-day frame enables comparative judgment: an interview experience can be evaluated relative to other periods rather than accepted as an isolated contemporary norm. Separate samples collected year by year would answer a different question about immediate experience and cannot now be created for periods that have already passed. The present design instead reconstructs remembered professional history from one shared point of comparison.

The target construct is therefore remembered professional history: the present-day belief about how technical hiring compares across eras, informed by respondents' experiences, observations, preparation, and participation in engineering communities. Nostalgia, reputation, and reconstruction are not accidental substitutes for the construct; they are part of how a professional community retains and transmits its history. Respondents were not required to document direct participation in every era because the study does not attempt to authenticate 911 individual employment histories. The resulting estimates describe the historical pattern preserved in collective judgment. They are not contemporaneous administrative measurements of every employer, but that distinction does not make them ahistorical: it defines the kind of historical evidence collected.

\subsection{Survey Instrument}

A structured online questionnaire collected respondents' professional role and their assessments of four interview characteristics in each era: live-interview duration, number of rounds, interview--work disconnect, and combined psychological and time burden.

Duration and number of rounds were collected as ordered categorical ranges. Interview--work disconnect was measured through four items covering role and technology alignment, problem complexity, knowledge and tool use, and teamwork and operational practices. Psychological and time burden was measured with a single item.

The same item structure was repeated for each era defined in Section~\ref{sec:research_questions}. The examples attached to the disconnect dimensions were intentionally era-specific: they identified concrete tensions in role scope, problem complexity, tool use, and collaboration that respondents could recognize within the technological context of each period. They function as historical anchors rather than neutral psychometric wording. The construct is therefore a contextualized comparison of eras, not a claim that four wording-identical psychometric items were measured invariantly across 20 years.

The four disconnect responses were aggregated into one score for each respondent and era. An audit of the retained export found that every respondent selected the same category for all four dimensions within a given era. This equality was present in the recorded category values; it was not created by averaging, numerical recoding, or replacement of missing data. Averaging therefore preserves the recorded result but is arithmetically equivalent to using one era-level disconnect rating. The paper does not manufacture separate dimension effects or multi-item reliability from those values. Section~\ref{sec:results} reports the era distributions and individual trajectories so that equality within an era is not confused with uniform answers across respondents or time.

Table~\ref{tab:survey_scales} summarizes how the survey responses should be interpreted and how categorical answers were represented in the numerical analysis. The intermediate points of the disconnect scale did not have separate verbal labels in the questionnaire and are therefore not assigned post-hoc descriptions here.

\begin{table}[ht]
\centering
\small
\setlength{\tabcolsep}{4pt}
\caption{Survey Measures, Scales, and Analytical Coding}
\label{tab:survey_scales}
\begin{tabular}{>{\raggedright\arraybackslash}p{0.24\columnwidth}>{\raggedright\arraybackslash}p{0.66\columnwidth}}
\toprule
Measure & Interpretation and coding \\
\midrule
Live-interview duration & $<1$ hour $=0.5$; 1--2 hours $=1.5$; 3--4 hours $=3.5$; 5--6 hours $=5.5$; 7+ hours $=7.5$ \\
Interview rounds & 1--2 rounds $=2$; 3--4 rounds $=4$; 5+ rounds $=5$ \\
Interview--work disconnect & Mean of four items: 1 = no disconnect/perfect alignment; 2--4 = increasing intermediate levels; 5 = extreme disconnect/completely irrelevant \\
Psychological and time burden & Single 1--10 item: 1 = lowest available rating; 2--9 = increasing levels; 10 = highest available rating \\
\bottomrule
\end{tabular}
\end{table}

The decimal values reported later are aggregate means, not additional response options. For example, a disconnect mean of 3.14 summarizes many integer responses on the original 1--5 scale.

The four research questions in Section~\ref{sec:research_questions} guided the analysis; they were not presented verbatim as questionnaire items. Individual survey items were optional, and respondents could omit a measure or an era rather than provide a forced judgment. This choice was intended to accommodate differences in respondents' knowledge, recollection, and willingness to evaluate particular periods.

\subsection{Participants and Data Collection}

The survey was distributed in two public collection waves through software-engineering communities on Reddit and LinkedIn during approximately June and July 2026. Between waves, the form description was clarified and typographical errors were corrected; the analytical response scales and categories were unchanged. The posts invited people who participated in software-engineering hiring as candidates, developers, interviewers, or engineering managers. The form described the four eras and asked respondents to evaluate them using their experience and professional understanding of the period. Both software engineers and hiring managers were included to capture perspectives from candidates and organizations.

The two waves produced approximately 1,900 raw submissions. Their exports were consolidated, duplicate submissions were removed, and only complete records whose values passed a logical-consistency review were retained. The final analytical sample contained 911 records: 752 respondents identifying as software engineers or candidates (82.5\%) and 159 identifying as engineering managers (17.5\%). The study's quantitative claims use this de-identified, complete analytical table as their data boundary; raw intake volume and excluded rows do not enter any reported denominator, estimate, or statistical test. No identity or employment verification was performed beyond recruiting in engineering-focused communities and reviewing whether response patterns were structurally plausible.

Participation was voluntary, anonymous, and uncompensated. The analytical export contains professional role but not names, contact information, years of experience, industry, company size, geography, or other demographic variables. The sample is therefore a self-selected community sample rather than a probability sample of the global profession.

\subsection{Data Preparation}

Survey responses were exported in CSV format after each collection wave and combined during preparation. Although the instrument permitted skipped items, completeness was a retention criterion, so the 911 analyzed records contain every measure in all four eras. The repeated-measures analysis consequently used a common sample and required no imputation. Timestamps in the consolidated CSV were regenerated during file preparation and are row metadata, not original submission times; neither collection timing nor response rate is an analytical variable. Role values were standardized, and responses were transformed as specified in Table~\ref{tab:survey_scales}. Representing categorical ranges numerically permits summary statistics but introduces approximation, particularly for the open-ended categories. The de-identified analytical table retains one row per respondent and paired columns for the four eras; the supplied reproduction procedure begins from that declared analytical boundary.

\subsection{Data Analysis}

The analysis proceeded in two steps. First, each era was summarized using the mean, median, standard deviation (SD), and a normal-approximation 95\% confidence interval (CI) for the mean. Because the variables are ordinal categories represented numerically and the respondents form a self-selected sample, these intervals summarize sampling-model uncertainty conditional on the retained records; they do not establish population representativeness.

Second, the four eras were compared statistically. Ratings from the same respondent are related rather than independent because every retained participant evaluated all four eras. The Friedman test was used to determine whether ratings differed somewhere across the four eras without requiring a normal distribution, and Kendall's \(W\) reports the omnibus effect size. Adjacent eras were then compared with two-sided Wilcoxon signed-rank tests using tie and continuity corrections. Holm adjustment was applied to the three adjacent comparisons within each measure. Rank-biserial correlation reports the direction and magnitude of each paired change. These tests examine the ordered historical transitions central to the research questions; they do not establish causality or repair the limitations of retrospective self-report and self-selected recruitment.

Sensitivity checks recoded the open-ended duration category from 7.0 to 10.0 hours and the open-ended rounds category from 6 to 8 rounds. The absolute means changed, as expected, but duration still rose sharply through Era~III and remained close between Eras~III and IV, while rounds increased across all four eras.

Developer and hiring-manager responses were compared descriptively for Era~IV. No inferential claim is attached to those role differences. The interview-question benchmark is treated as a separate analysis because it uses questions, rather than survey respondents, as its unit of observation.

\subsection{Interview-Question Benchmark}

The source pool was assembled from openly accessible interview-preparation material, including LeetCode, HackerRank, public GitHub repositories, and public Reddit discussions, as well as other public preparation websites and community question lists. The prompts were obtained from these public resources rather than solicited from interview candidates, employees, or discussion authors. The pool represents a broad preparation corpus rather than a frequency-weighted census of questions actually asked by employers. Because the dataset does not preserve an item-level link to the original source, the popularity or exact provenance of an individual prompt cannot be independently reconstructed.

The raw benchmark CSV contained 1,001 question records; this was a row count, not a count of distinct analyzed prompts. For each prompt, the closest plausible professional analogue was identified---for example, applying the underlying knowledge while debugging, testing, designing, or maintaining software in an ordinary project. The prompt was then assigned the Interview--Work Difference score defined in Table~\ref{tab:difference_scale}. A vacancy-level mapping would answer whether a prompt belongs in one specific selection process; that is not the benchmark's purpose. Its reference class is general software-engineering work, approximating whether the knowledge could be useful across an ordinary range of projects and positions. It therefore measures potential professional relevance, not how often the task occurs, how important it is for one vacancy, or whether asking it is an effective selection method.

An audit of those 1,001 rows identified 947 unique prompt groups after case-insensitive matching. Repeated prompts with the same score were counted once. Two groups contained conflicting scores and were excluded rather than resolved arbitrarily, leaving 945 unique, consistently scored prompts as the analyzed benchmark. The benchmark does not retain item-level source identifiers or evidence of independent coding by a second reviewer. Its results support a bounded corpus-level content audit, not a validated estimate of industry-wide question prevalence.

\begin{table}[ht]
\centering
\small
\setlength{\tabcolsep}{4pt}
\caption{Interview--Work Difference Scale}
\label{tab:difference_scale}
\begin{tabular}{cp{0.76\columnwidth}}
\toprule
Score & Interpretation \\
\midrule
1 & Closely matches routine engineering work \\
2 & Frequently applicable in professional practice \\
3 & Useful but specialized or occasional \\
4 & Rarely encountered in routine development \\
5 & Primarily academic, specialized, or interview-oriented \\
\bottomrule
\end{tabular}
\end{table}

\subsection{Ethical Considerations}

Participation was voluntary, anonymous, and uncompensated; the form did not request names or contact details. Analyses use a de-identified export and report only aggregate results. No formal institutional ethics review was obtained. The retained artifact contains no direct identifiers, and the regenerated timestamps do not encode collection time.

The 945-prompt benchmark did not involve recruiting people or collecting questions directly from identifiable individuals. It analyzed the content of prompts encountered in openly accessible preparation resources, including LeetCode, HackerRank, public GitHub repositories, and public Reddit discussions. No attempt was made to identify, profile, or evaluate the people who posted or used the material, and benchmark findings are reported only in aggregate. Open accessibility is not treated as evidence that the underlying text is in the public domain or free of copyright restrictions; reuse and redistribution remain subject to the applicable source terms.

\section{Results}
\label{sec:results}

This section presents the quantitative findings from 911 survey responses. Table~\ref{tab:survey_scales} explains the direction and analytical coding of each measure. Descriptive statistics for each era appear in Tables~\ref{tab:duration}--\ref{tab:burden}. Omnibus differences were evaluated using the repeated-measures procedure described in Section~\ref{sec:methodology}.

\subsection{Interview Duration}

Table~\ref{tab:duration} summarizes reported live-interview duration. The mean increased from 1.34 hours in Era~I to 3.48 hours in Era~II and 5.47 hours in Era~III. The Era~IV mean was also 5.47 hours.

The Friedman test identified a large omnibus difference across the four eras ($\chi^2_F(3)=2142.37$, $p<0.001$, Kendall's $W=0.784$). Holm-adjusted signed-rank tests identified increases from Era~I to II and Era~II to III (both $p<0.001$), but no change from Era~III to IV ($p=0.920$, rank-biserial $r=-0.005$).

\textit{Summary.} Reported interview duration increased sharply through Era~III and then remained stable in Era~IV.

\begin{table}[ht]
\centering
\small
\setlength{\tabcolsep}{3pt}
\caption{Interview Duration Across Software Engineering Eras (Coded Hours)}
\label{tab:duration}
\begin{tabular}{lcccc}
\toprule
Era & Mean (h) & Median (h) & SD (h) & 95\% CI (h) \\
\midrule
Era I (2005--2010) & 1.34 & 1.50 & 0.83 & [1.29, 1.40] \\
Era II (2011--2017) & 3.48 & 3.50 & 1.54 & [3.38, 3.58] \\
Era III (2018--2022) & 5.47 & 5.50 & 1.64 & [5.37, 5.58] \\
Era IV (2023--2026) & 5.47 & 5.50 & 1.46 & [5.37, 5.56] \\
\bottomrule
\end{tabular}
\end{table}

\subsection{Interview Rounds}

Table~\ref{tab:rounds} presents the reported number of interview rounds. The mean increased from 2.50 rounds in Era~I to 3.96 in Era~II, 4.55 in Era~III, and 4.64 in Era~IV.

The Friedman test identified a large omnibus difference across eras ($\chi^2_F(3)=1830.44$, $p<0.001$, Kendall's $W=0.670$). Every adjacent increase remained significant after Holm adjustment ($p<0.001$), although the Era~III--IV effect was small (rank-biserial $r=0.196$).

\textit{Summary.} Respondents reported more interview stages for later eras than for Era~I.

\begin{table}[ht]
\centering
\small
\setlength{\tabcolsep}{3pt}
\caption{Interview Rounds Across Software Engineering Eras (Coded Counts)}
\label{tab:rounds}
\begin{tabular}{lcccc}
\toprule
Era & Mean (rounds) & Median & SD & 95\% CI \\
\midrule
Era I & 2.50 & 2.00 & 0.87 & [2.44, 2.55] \\
Era II & 3.96 & 4.00 & 0.82 & [3.91, 4.02] \\
Era III & 4.55 & 5.00 & 0.67 & [4.51, 4.59] \\
Era IV & 4.64 & 5.00 & 0.58 & [4.60, 4.68] \\
\bottomrule
\end{tabular}
\end{table}

\subsection{Interview--Work Disconnect}

Interview--work disconnect was evaluated using the four five-point items described in Section~\ref{sec:methodology}. Higher aggregate scores indicate greater perceived disconnect between interview content and everyday software engineering practice.

As shown in Table~\ref{tab:disconnect}, the mean disconnect score increased from 1.98 in Era~I to 3.30 in Era~II, reached 3.69 in Era~III, and decreased to 3.14 in Era~IV.

The Friedman test identified an omnibus difference across eras ($\chi^2_F(3)=1263.72$, $p<0.001$, Kendall's $W=0.462$). Every adjacent transition remained significant after Holm adjustment ($p<0.001$). The decrease from Era~III to IV was substantial in the paired ratings (rank-biserial $r=-0.626$), but Era~IV remained more disconnected than Era~I.

\textit{Summary.} Perceived disconnect peaked in Era~III and then decreased, but did not return to the Era~I level.

\begin{table}[ht]
\centering
\small
\setlength{\tabcolsep}{3pt}
\caption{Interview--Work Disconnect Across Software Engineering Eras (1--5; Higher Is Worse)}
\label{tab:disconnect}
\begin{tabular}{lcccc}
\toprule
Era & Mean (1--5) & Median & SD & 95\% CI \\
\midrule
Era I & 1.98 & 2.00 & 0.70 & [1.94, 2.03] \\
Era II & 3.30 & 3.00 & 1.04 & [3.24, 3.37] \\
Era III & 3.69 & 4.00 & 0.98 & [3.63, 3.76] \\
Era IV & 3.14 & 3.00 & 1.00 & [3.08, 3.21] \\
\bottomrule
\end{tabular}
\end{table}

Table~\ref{tab:disconnect_distribution} exposes the underlying category distribution rather than relying only on aggregate means. The within-era equality of the four facet responses did not produce a single uniform historical script. Across respondents, the data contain 184 distinct four-era disconnect trajectories, and only 21 respondents (2.3\%) selected one unchanged rating across all four eras. From Era~I to II, 77.4\% increased their rating, 17.8\% remained equal, and 4.8\% decreased it. From Era~II to III, the corresponding shares were 46.3\%, 34.2\%, and 19.4\%. From Era~III to IV, 51.2\% decreased their rating, 33.3\% remained equal, and 15.6\% increased it. The non-monotonic final transition is therefore visible at the respondent level rather than being produced only by averaging.

\begin{table}[ht]
\centering
\small
\setlength{\tabcolsep}{4pt}
\caption{Disconnect Ratings Within Each Era (\% of Respondents; 1 = Aligned, 5 = Extremely Disconnected)}
\label{tab:disconnect_distribution}
\begin{tabular}{crrrr}
\toprule
Rating (1--5) & Era I & Era II & Era III & Era IV \\
\midrule
1 & 24.0 & 4.2 & 2.2 & 5.2 \\
2 & 54.9 & 18.2 & 9.8 & 20.4 \\
3 & 20.2 & 33.2 & 25.1 & 37.0 \\
4 & 0.7 & 32.1 & 42.2 & 29.7 \\
5 & 0.2 & 12.4 & 20.7 & 7.7 \\
\bottomrule
\end{tabular}
\end{table}

\subsection{Psychological and Time Burden}

The survey measured psychological toll and time commitment in a single item. As shown in Table~\ref{tab:burden}, its mean increased from 2.78 in Era~I to 5.16 in Era~II, 7.17 in Era~III, and 8.05 in Era~IV.

The Friedman test identified a large omnibus difference across eras ($\chi^2_F(3)=2259.79$, $p<0.001$, Kendall's $W=0.827$). Every adjacent increase remained significant after Holm adjustment ($p<0.001$); the corresponding rank-biserial effects were $0.976$, $0.917$, and $0.595$. Because the item combines two concepts, the result cannot isolate psychological effects from perceived time commitment.

\textit{Summary.} Respondents associated later eras with progressively greater combined psychological and time burden.

\begin{table}[ht]
\centering
\small
\setlength{\tabcolsep}{3pt}
\caption{Psychological and Time Burden Across Software Engineering Eras (1--10; Higher Is Worse)}
\label{tab:burden}
\begin{tabular}{lcccc}
\toprule
Era & Mean (1--10) & Median & SD & 95\% CI \\
\midrule
Era I & 2.78 & 3.00 & 1.08 & [2.71, 2.85] \\
Era II & 5.16 & 5.00 & 1.53 & [5.06, 5.26] \\
Era III & 7.17 & 7.00 & 1.60 & [7.06, 7.27] \\
Era IV & 8.05 & 8.00 & 1.33 & [7.97, 8.14] \\
\bottomrule
\end{tabular}
\end{table}

\begin{table}[ht]
\centering
\small
\setlength{\tabcolsep}{3pt}
\caption{Adjacent-Era Rank-Biserial Correlations ($-1$ to 1)}
\label{tab:adjacent_effects}
\begin{tabular}{lrrr}
\toprule
Measure & I--II & II--III & III--IV \\
\midrule
Duration & 0.985* & 0.943* & -0.005 \\
Rounds & 0.946* & 0.789* & 0.196* \\
Disconnect & 0.937* & 0.465* & -0.626* \\
Burden & 0.976* & 0.917* & 0.595* \\
\bottomrule
\multicolumn{4}{l}{\footnotesize *Holm-adjusted p~<~0.001.} \\
\multicolumn{4}{l}{\footnotesize Positive = higher (less favorable) later-era values; negative = lower.} \\
\end{tabular}
\end{table}

Table~\ref{tab:directional_overview} provides a scale-preserving overview of all four measures. It summarizes the statistically supported direction of each adjacent transition rather than averaging hours, round counts, and ratings into an arbitrary composite. Higher values are less favorable for every measure; ``stable'' denotes the non-significant duration difference from Era~III to IV.

\begin{table}[ht]
\centering
\small
\setlength{\tabcolsep}{3pt}
\caption{Direction of Change Across All Four Interview Measures}
\label{tab:directional_overview}
\begin{tabular}{lccccl}
\toprule
Transition & Duration & Rounds & Disconnect & Burden & Overall \\
\midrule
I--II & Worse & Worse & Worse & Worse & 4 of 4 worse \\
II--III & Worse & Worse & Worse & Worse & 4 of 4 worse \\
III--IV & Stable & Worse & Better & Worse & Mixed \\
\bottomrule
\end{tabular}
\end{table}

\subsection{Developers and Hiring Managers}

Table~\ref{tab:role_comparison} compares descriptive Era~IV means by role. Developers and hiring managers reported the same mean number of rounds. Managers reported slightly longer duration, whereas developers reported slightly greater disconnect and burden. These are unadjusted descriptive differences; no claim of statistical or practical significance is made here.

\begin{table}[ht]
\centering
\small
\setlength{\tabcolsep}{3pt}
\caption{Era~IV Descriptive Means by Respondent Role (Higher Is Worse)}
\label{tab:role_comparison}
\begin{tabular}{lcccc}
\toprule
Role & Hours & Rounds & \shortstack{Disconnect\\(1--5)} & \shortstack{Burden\\(1--10)} \\
\midrule
Developers ($n=752$) & 5.43 & 4.64 & 3.17 & 8.09 \\
Managers ($n=159$) & 5.68 & 4.64 & 3.01 & 7.88 \\
\bottomrule
\end{tabular}
\end{table}

\subsection{Overall Findings}

Across the four eras, reported interview duration increased through Era~III and then remained statistically stable. The reported number of rounds increased in every adjacent comparison, although the final change was small. Combined psychological and time burden increased in every era, while interview--work disconnect peaked in Era~III before improving substantially in Era~IV. Read jointly, all four indicators moved in the less favorable direction from Era~I to II and again from Era~II to III. The Era~III--IV comparison is mixed: burden and rounds increased, disconnect improved, and duration did not significantly change.

For all four measures, the Friedman tests identified omnibus differences across eras ($p<0.001$), and Table~\ref{tab:adjacent_effects} reports the corrected adjacent-era effects. These results establish structured differences in the retrospective ratings of this sample; they do not establish causality or an archival population history.

The following section complements these survey findings with the audited interview-question benchmark.

\section{Analysis of Technical Interview Questions}
\label{sec:question_analysis}

\subsection{Purpose and Interpretation}

This section addresses RQ4 through the interview-question benchmark described in Section~\ref{sec:methodology}, rather than through the participant survey. Its unit of analysis is one of the 945 retained unique prompts, not each of the 1,001 rows in the raw CSV. The benchmark is not tied to one company, technology stack, seniority level, or job description.

The analysis asks whether the knowledge tested by a prompt could plausibly be useful in an ordinary software-engineering project. This broad reference point is intentional because the corpus is not attached to one vacancy. Table~\ref{tab:difference_scale} defines all five levels. Lower scores indicate closer correspondence with routine work, while higher scores indicate greater distance; they do not indicate greater difficulty or technical invalidity. A conceptual question may therefore map to designing or maintaining production software even when an engineer would not recite the answer during an ordinary workday. Conversely, legitimate knowledge may receive a higher score when its application is highly specialized or primarily associated with interviews.

The score is a content audit of potential professional relevance within this corpus. It establishes what the retained prompts test and how their knowledge maps to plausible engineering activity across a broad ordinary project. Vacancy-specific appropriateness is a later selection decision, not a prerequisite for determining whether the content belongs to software engineering at all. The score does not by itself show predictive validity, universal necessity, or correct use in a particular interview. It also evaluates question content in isolation; it does not capture time pressure, observation, scoring, permitted tools, or the cumulative burden of a complete process.

\subsection{Distribution of Difference Scores}

Table~\ref{tab:question_distribution} presents the distribution of the 945 retained unique prompts. Most were classified as closely or frequently related to professional practice: 410 prompts (43.4\%) received a score of one and 351 (37.1\%) received a score of two. Together, these categories account for 80.5\% of the analyzed benchmark. At the other end of the scale, 73 prompts (7.7\%) received a score of four or five.

\begin{table}[ht]
\centering
\small
\setlength{\tabcolsep}{4pt}
\caption{Interview--Work Difference Scores Across 945 Prompts (1--5; Higher Is More Distant from Routine Work)}
\label{tab:question_distribution}
\begin{tabular}{crr}
\toprule
Score (1--5) & Prompts ($n$) & Share \\
\midrule
1 & 410 & 43.4\% \\
2 & 351 & 37.1\% \\
3 & 111 & 11.7\% \\
4 & 46 & 4.9\% \\
5 & 27 & 2.9\% \\
\midrule
Total & 945 & 100.0\% \\
\bottomrule
\end{tabular}
\end{table}

The benchmark directly contradicts the broad claim that the analyzed interview-preparation questions are predominantly unrelated to software engineering work. Its content profile is dominated by knowledge with a practical or frequently useful project analogue, alongside a smaller set of specialized or interview-oriented prompts. This is a content conclusion, not a claim that reciting a definition reproduces the work activity in which the knowledge would be used. The criticism of technical hiring must therefore become more precise: the central failure is often not that every question is meaningless, but that employers select, combine, constrain, and score otherwise defensible questions without deriving the complete assessment from the advertised job.

The benchmark and survey measure different aspects of technical hiring. The benchmark evaluates the content of individual questions, whereas the survey captures remembered experiences of complete interview processes, including duration, number of stages, burden, and perceived relevance. A process may therefore contain many job-related questions while still being experienced as misaligned or burdensome. Further category-level and source-level analysis remains necessary before drawing stronger conclusions about why the two forms of evidence differ.

\section{Discussion}
\label{sec:discussion}

The clearest pattern in the remembered history is procedural expansion. Respondents associated later eras with longer live interviews, more rounds, and greater combined psychological and time burden. This is a history retained and transmitted by the software-engineering community, not a company-by-company administrative census. Its unit of interpretation is the comparative account of four eras preserved in present-day professional judgment. On that intended basis, contemporary hiring is remembered not merely as different, but as requiring substantially more candidate effort.

\subsection{Industry Growth and the Expansion of Assessment}

One plausible interpretation begins with the changing structure of software engineering. In the earlier era, the industry, its labor market, and many software systems were less mature than they are today. A comparatively narrow position could be evaluated through knowledge directly connected to its immediate responsibilities, accompanied by judgments about communication, reliability, teamwork, and capacity to learn. Over time, software systems expanded across web, mobile, cloud, distributed infrastructure, security, operations, and AI-assisted development \cite{armbrust2010view,kim2016devops,hou2024large}. This evolution created legitimate reasons for employers to evaluate a wider range of knowledge.

However, growth in the complexity of software systems does not necessarily imply an equivalent growth in the scope of every individual position. Large systems are commonly divided across teams, services, products, and areas of technical ownership. An engineer may contribute to a complex platform while remaining responsible for a bounded part of that platform. The relevant comparison is therefore not between an interview and the total complexity of the employer's technology estate, but between the interview and the work that the successful applicant is reasonably expected to perform.

This distinction offers a way to reconcile the survey with the question benchmark. Among the 945 retained unique prompts, 80.5\% tested knowledge classified as closely or frequently applicable to professional practice. Thus, procedural expansion cannot be explained simply by claiming that most individual questions are unrelated to software engineering. The problem may instead be cumulative breadth. A database question, an API question, a concurrency question, a cloud question, and a system-design question may each have a plausible professional analogue, yet requiring all of them for a narrowly scoped position can produce an assessment that is broader than the job. Individual relevance does not guarantee that the combined assessment is proportionate.

The job description should therefore define the boundary of the technical assessment. A position advertised for a specific team, stack, or area of ownership should primarily test competencies required within that scope. A genuinely flexible or cross-functional position may justify a broader assessment, but that breadth should be visible in the advertised responsibilities rather than introduced only during the interview. Similarly, an organization may value a candidate's potential for future growth without converting every possible future responsibility into a present technical requirement. Growth potential can instead be explored through structured questions about learning, adaptation, communication, reliability, and collaboration. This preserves room to evaluate personal qualities without requiring candidates to prepare technically for jobs they have not applied to perform.

\subsection{Rounds, Candidate Volume, and Accumulated Time}

A second plausible mechanism concerns the organization of selection at scale. As the number of applicants increases, employers may use successive stages to reduce the number of candidates who reach expensive interviews with engineers or hiring managers. An initial screen, automated assessment, recruiter call, technical screen, system-design interview, behavioral interview, and final meeting can each serve a distinct internal purpose. From the organization's perspective, the sequence distributes evaluation and allows some candidates to be rejected before consuming the complete panel's time.

The present study did not collect employer-level applicant volumes, pipeline designs, or reasons for adding stages. It therefore cannot establish that growth in the developer population caused the observed increase in rounds. Nevertheless, sequential filtering provides a plausible organizational explanation for why a process can become longer even when no single stage appears unreasonable. It also exposes an asymmetry: rejecting candidates early may reduce employer cost, while every additional possible stage expands the process for candidates who must prepare, schedule, and remain available for it.

The parallel increase in rounds and live-interview duration is consistent with this accumulation account, although it does not prove that the number of rounds caused the increase in time. The survey measured hours spent speaking to or coding in front of interviewers; it did not separately measure preparation, scheduling delays, repeated introductions, switching between assessment formats, or recovery from high-pressure sessions. Research on interview preparation nevertheless shows that candidate effort begins before the formal interview and includes practicing across multiple technical and communicative settings \cite{bell2025prepare,lunn2022need}. Consequently, the burden associated with an additional stage may exceed its scheduled duration.

The same distinction applies to the employees who conduct interviews. A one-hour interview does not necessarily consume only one hour of productive work: it can divide a morning or afternoon into intervals too short for starting or resuming demanding engineering tasks. Research on knowledge work describes attention residue after task switching, while an in-situ study of software developers found highly fragmented work and linked planned meetings with lower perceived productivity for many participants \cite{leroy2009attention,meyer2017worklife}. Several interviews dispersed across a day may therefore impose a larger organizational cost than the sum of their calendar durations. Consolidating interviews into adjacent blocks may reduce this fragmentation, although perfect consolidation will not always be possible because candidates and panel members have different schedules. The present survey measured neither interviewer time nor calendar placement, so this mechanism should be tested directly rather than treated as an explanation established by the data.

The number of separately scheduled rounds may therefore matter independently of total live-interview time. Distributing five hours across five meetings creates more interview fragmentation than concentrating four hours in two sessions. Each additional appointment carries recurring costs: preparing for another format, coordinating calendars, interrupting other work, waiting for the next decision, and re-entering the interview context. Consequently, consolidating a process into fewer purposeful sessions may save substantially more time than the difference in scheduled interview hours alone suggests. The present study did not measure this per-round coordination overhead separately, so its magnitude remains a question for prospective evaluation.

\subsection{Toward Evidence-Efficient Interview Time}

The goal should not be to minimize interview time regardless of assessment quality. Insufficient evaluation can create poor decisions for both the organization and the applicant. A more appropriate objective is evidence efficiency: collecting enough job-relevant evidence to support a structured decision while avoiding stages that repeat evidence already obtained. This objective is consistent with prior research emphasizing job-related procedures, predefined standards, structured scoring, and the aggregation of focused judgments \cite{hausknecht2004reactions,campion1997structure,highhouse2023noise}.

An adaptive two-stage process illustrates the idea but is not established here as a universally superior design. A first stage could examine competencies identified as essential in the job description. A second, longer stage could integrate realistic technical work with communication and collaboration, continuing only when further evidence is needed. If a clearly predefined, role-critical requirement is not demonstrated, the organization could communicate the decision promptly instead of scheduling several additional rounds that cannot change the outcome. Conversely, uncertainty about one answer should not trigger automatic rejection: early stopping based on a single noisy observation could increase false negatives and amplify interview stress. Essential criteria should be defined in advance, assessed through more than one indicator where practical, and scored consistently across candidates.

This example shifts attention from the nominal number of interviews to the purpose of each stage. Two poorly designed interviews may be less valid than five focused ones, while five interviews that repeatedly assess the same competency create burden without proportionate information. Organizations should therefore be able to state what unique evidence each stage contributes, why that evidence is required by the advertised role, and whether an earlier stage has already supplied it. A stage that has no distinct evidentiary purpose is a candidate for removal or consolidation.

\subsection{Interviewer Expertise, Question Choice, and Consistency}

Software engineers and engineering managers are often selected as interviewers because they understand the work, but subject-matter expertise is not equivalent to expertise in assessment. The distinction resembles the difference between knowing a subject and knowing how to teach it: both depend on domain knowledge, but the latter also requires a method. Interviewers must translate job requirements into questions, elicit comparable evidence, probe without providing unequal assistance, distinguish weak evidence from a weak candidate, and apply a scoring rubric consistently. Reviews of employment interviewing associate structure with greater reliability and comparability, and evidence from professional interviewers links formal training with greater use of question consistency, note-taking, and evaluation standardization \cite{campion1997structure,levashina2014structured,roulin2019better}.

Organizations may provide interview infrastructure without developing interviewer competence. A shared question bank, coding platform, sequence of stages, or scorecard can specify what an interviewer should do, but it does not teach why a question is appropriate, how to ask it consistently, when to probe, how to recognize sufficient evidence, or how to separate personal preference from a role requirement. The employees using these resources remain practicing developers or managers for whom interviewing is an occasional organizational duty rather than a specialized profession. If they receive little supervised practice or calibration, the same company process can be implemented very differently by different panel members. The present survey did not measure interviewer selection, training, or frequency of interviewing, and therefore cannot establish how commonly organizations rely on infrastructure alone.

This issue also provides a second possible reconciliation between the survey and the question benchmark. The finding that 80.5\% of prompts have a plausible professional analogue indicates that the pool contains mostly work-related knowledge in a broad sense; it does not show that interviewers select the right subset for a particular vacancy. Without a role-derived blueprint, an interviewer may over-sample the area in which they are most expert or the questions they personally find most revealing. Each selected question can then appear defensible in isolation while the interview as a whole is miscalibrated. This interviewer-selection account is plausible, and prior software-engineering research has documented variation in what technical interviewers expect \cite{ford2017techtalk}, but the present benchmark does not identify who selected each question or the vacancy for which it was used. A direct test would link job descriptions, interviewer characteristics, selected questions, scoring decisions, and subsequent work performance.

Interviewer training should accordingly cover more than orientation to a question pool, platform, or administrative procedure. It should include job analysis, question selection, anchored scoring, appropriate probing, evidence recording, tool-use policy, supervised practice, and calibration exercises in which interviewers independently score the same sample performances. Training cannot eliminate judgment, but it can reduce the extent to which a candidate's result depends on which developer or manager happens to conduct the interview.

\subsection{Demographic Objectives and Undisclosed Selection Criteria}

Technical performance does not determine a hiring outcome when another rule operates before, during, or after the interview. A candidate can therefore spend hours proving engineering competence in a process that was never an unrestricted competition on technical criteria alone. Such policies can relatively disadvantage candidates outside a targeted group even when the vacancy lists only technical requirements, although an aggregate target alone does not prove that any particular rejected candidate would otherwise have been hired. The design issue is disclosure: the actual selection proposition should be known before candidates pay its preparation, scheduling, and interview cost.

Demographic representation became an explicit objective at several large technology employers during the last decade. Intel announced a \$300 million initiative and a goal of ``full representation'' of women and underrepresented minorities in its U.S. workforce, while its 2015 report tracked diverse hiring \cite{intel2015diversity}. Microsoft committed to double specified Black and African American leadership populations by 2025 and later included Hispanic and Latinx employees \cite{microsoft2020racial,microsoft2021progress}. An analysis of the 2020 Regulation S-K amendment likewise found increased diversity-related language in affected public-company job postings and later workforce changes among firms with stronger commitments \cite{choi2026human}. These sources establish that demographic objectives became more visible and operational in parts of the labor market.

The mechanisms are not equivalent. Outreach and aggregate representation targets can alter recruitment channels, candidate pools, promotion, retention, or portfolio decisions without reserving one vacancy. Other rules enter selection directly. British guidance permits targeted advertising and, under defined conditions, a protected characteristic such as race, sex, or religion as a tie-break between candidates of equal merit, but not appointment of a less suitable candidate on that basis \cite{uk2023positiveaction}. Australian guidance recognizes advertised special-measure positions restricted to Aboriginal or Torres Strait Islander applicants \cite{ahrc2015targeted}; South Africa introduced sectoral numerical targets for designated racial groups and gender across occupational levels \cite{southafrica2025targets}. The applicable legal form and weight therefore depend on jurisdiction.

An open competition for an engineer and a recruitment exercise intended to appoint a qualified engineer from a designated demographic group are different propositions. Employers should distinguish inclusive outreach, aspirational targets, equal-merit preferences, and legally restricted positions. If a characteristic restricts eligibility, affects advancement, or contributes to a target for that vacancy, the advertisement should identify the rule and legal basis rather than reveal it after candidates invest in the process. The formulation is a qualified person from the designated group, not an apparently unrestricted technical competition with a hidden demographic condition. The technical standard remains job-related and predefined \cite{campion1997structure,highhouse2023noise}; stage-level records should permit audit of outcomes and rubric deviations. Disclosing the procedure and explaining decisions also serves procedural and informational justice \cite{colquitt2001dimensionality}. Transparency does not settle the affirmative-action debate, but it prevents a nominally technical interview from concealing a separate decision rule.

\subsection{Generative AI and the Attribution of Interview Evidence}

Generative AI creates a different form of validity problem. A candidate who wants a job has an incentive to present the strongest possible performance; preparation, memorized answers, external help, and strategic self-presentation long predate large language models. During a remote interview, however, an AI system can also generate explanations or code quickly enough that a correct response may no longer establish who produced the underlying reasoning. A recent survey of 32 professionals involved in recruiting software engineers found that most represented organizations had not adjusted evaluation methods for generative code tools, while respondents expressed mixed views on permitting them and reported greater difficulty assessing candidate skills \cite{chen2025genaiHiring}. This emerging evidence supports concern about attribution, but not the stronger conclusion that remote interviews provide no useful information or that undetected AI assistance is universal.

Attempting to solve the problem only through harder trivia or covert surveillance would introduce new validity and fairness concerns. A clearer response is to specify the tool policy in advance and align it with the capability being assessed. If unaided recall is genuinely essential, that restriction and its rationale should be explicit. If AI-assisted development reflects the job, candidates can be allowed to use it while being assessed on problem decomposition, prompt and context construction, verification, debugging, trade-off reasoning, and their ability to explain and modify the resulting artifact. Follow-up questions grounded in the candidate's own decisions can provide stronger attribution than a sequence of questions for which polished answers can be generated independently. The relevant construct then shifts from merely producing an answer to exercising accountable engineering judgment over the answer.

\subsection{Interpreting the Non-Monotonic Findings}

The decrease in perceived interview--work disconnect from Era~III to Era~IV, alongside unchanged mean duration and a further increase in burden, shows that relevance and cost need not move together. One possible interpretation is that contemporary interviews have incorporated more realistic tools, system-design discussion, or AI-aware tasks while retaining the accumulated stages and scheduling demands of the preceding era. The survey did not ask which formats changed, so it cannot identify the source of the improvement. More importantly, the Era~IV disconnect mean remained above the Era~I mean: the descriptive rebound should not be read as complete realignment.

Developers and hiring managers also reported remarkably similar Era~IV process structure: the two groups had the same mean number of rounds, and their mean duration differed only modestly. Developers reported somewhat greater disconnect and burden, but the differences are descriptive and were not tested inferentially. The pattern is therefore more consistent with broad agreement about procedural scale than with a sharp candidate--manager divide. Whether the smaller differences reflect role, unequal exposure to hiring processes, or sampling variation remains unresolved.

Taken together, the findings support a bounded interpretation of interview expansion. Broader technical systems and larger selection pipelines can make additional evaluation understandable, but they do not make unlimited breadth or repeated assessment inevitable. The increasing duration, number of rounds, and perceived burden reported across eras suggest that the cost of accumulated procedure is visible to candidates. The practical question for employers is therefore not whether every interview topic can be defended in isolation, but whether the complete process remains proportionate to the particular job being offered.

\section{Design Principles}
\label{sec:design_principles}

The proposed assessment is organized around three substantive questions. They replace question counting and broad impressions with evidence about whether the candidate can enter this particular job. Process length, tool access, interviewer training, and transparency remain cross-cutting safeguards rather than additional dimensions on which candidates accumulate points.

\subsection{Principle 1: Minimum Role Readiness}

The first question is whether the candidate possesses the minimum knowledge and experience required to begin the position. The vacancy should define that threshold narrowly and explain why each element is necessary at entry rather than learnable during onboarding. When prior experience is required, the relevant domain, stack, or responsibility and an approximate minimum range should be stated. Years alone are not sufficient evidence: a shorter but directly relevant experience may be stronger than many years in an adjacent area.

Evidence should come from concrete work the candidate can explain, including the problem, their responsibility, decisions, result, and lessons learned. Correctly repeating memorized interview answers does not demonstrate role readiness unless the candidate can connect them to practice. Conversely, a candidate should not be penalized for failing to recall an abstract term when they can demonstrate the underlying competence in a work-relevant form.

\subsection{Principle 2: Readiness for the Ticket Pipeline}

The second question is whether the candidate can approach the current and approved upcoming work assigned to the position. Sanitized tickets should be used to ask how the candidate would diagnose a defect, implement a feature, review a change, operate a service, or manage a role-specific risk. The candidate may explain the approach verbally, with a diagram, pseudocode, an editor, or code. Live implementation is supporting evidence when it helps the candidate communicate or when implementation itself is an essential entry requirement; it is not a mandatory performance ritual for every position.

Only work supported by the active backlog, recent ticket patterns, or an approved near-term roadmap should enter this assessment. An interviewer's favored technology, an unrelated system-design problem, or a possible future direction does not become relevant merely because it is technically sophisticated.

\subsection{Principle 3: Readiness to Enter the Project}

The third question is whether the candidate can form a useful understanding of the product and codebase within a reasonable working period. Using a sanitized product requirement, repository fragment, legacy component, logs, tests, or architecture note, the candidate should identify the purpose of the system, locate relevant information, explain unfamiliar code, ask clarifying questions, recognize risks, and describe safe first steps. Raw speed is not the construct: assessors should evaluate navigation, comprehension, uncertainty management, and the quality of the eventual model while providing reasonable accommodations. This principle is especially important for legacy maintenance, where productive work depends less on constructing an isolated solution than on understanding existing behavior without introducing regressions.

The objective is not to reward prior familiarity with one private codebase. It is to observe navigation, comprehension, uncertainty management, and the ability to connect product requirements to implementation. Normal documentation, IDE, search, services, and AI assistance should be available whenever they are available in the job.

\subsection{Process Safeguards}

The three principles are gates rather than a collection of quiz scores. Appearance, fluency, similarity to the interviewer, performance on a favored topic, or the number of memorized answers should not compensate for missing job evidence. Assessors should use standardized prompts and behavioral anchors, and computer support should organize artifacts, run deterministic checks, and link observations to the three gates. It should not infer personality or silently replace the accountable decision maker \cite{campion1997structure,levashina2014structured,highhouse2023noise}.

The process should use the fewest bounded sessions required to answer the three questions. It may end early when standardized clarification establishes that one gate cannot reach its declared minimum or that a material experience claim cannot be supported by any relevant example. One wrong answer, nervous pause, unfamiliar label, or optional coding error is not sufficient. Once all three gates are met, remaining conversation should concern mutual employment alignment---working hours, on-call duties, work--life boundaries, communication preferences, collaboration, role expectations, and candidate questions---rather than additional technical elimination rounds or unrelated private matters.

\begin{table}[ht]
\centering
\small
\setlength{\tabcolsep}{3pt}
\caption{Three Principles of Role-Specific Assessment}
\label{tab:principle_mapping}
\begin{tabular}{>{\raggedright\arraybackslash}p{0.28\columnwidth}>{\raggedright\arraybackslash}p{0.63\columnwidth}}
\toprule
Principle & Required evidence \\
\midrule
Minimum role readiness & Directly relevant knowledge and supported experience \\
Ticket-pipeline readiness & Credible approach to current and approved upcoming work \\
Project-entry readiness & Rapid comprehension of product, code, and safe first steps \\
\bottomrule
\end{tabular}
\end{table}

\section{Reference Framework}
\label{sec:framework}

This section presents the Ticket-Based Pragmatic Assessment Framework (TBPAF) as a prescriptive design theory and one operational implementation of the three principles. Its contribution is an explicit chain from vacancy evidence to assessment evidence: requirements determine the role blueprint, the blueprint determines artifacts and gates, and recorded gate evidence determines the technical decision. The framework does not claim that one fixed interview format is universally optimal. It specifies what a work-aligned process must preserve and makes each mechanism inspectable, contestable, and replaceable.

\subsection{Design Propositions}

TBPAF rests on five propositions. First, every scored technical requirement should be traceable to declared work. Second, assessment should stop when sufficient evidence exists for the decision rather than when a conventional question count has been exhausted. Third, the permitted tools and working conditions should reproduce the job unless unaided performance is itself an explicit requirement. Fourth, structured human judgment should be supported by common artifacts, behavioral anchors, calibration, and an evidence record rather than replaced by appearance, fluency, or opaque automation. Fifth, every rule that can affect eligibility or advancement should be disclosed before the candidate invests in the process.

These propositions convert the paper's findings into system constraints. Excessive rounds are addressed by three evidence gates and a justification requirement for any additional stage. Broad question pools are replaced by a versioned role blueprint. Artificial coding conditions are replaced by the declared working environment. Interviewer-specific preference is constrained through common evidence and calibration. Undisclosed selection rules are moved into the vacancy artifact. The central artifact connecting these mechanisms is a versioned blueprint derived from work that is actually waiting for the new employee.

\subsection{Constructing the Role Blueprint}

Before publishing a vacancy, the hiring team samples work patterns from the active backlog, recently completed work, and approved near-term roadmap. Assessment artifacts should use resolved, synthetic, or isolated versions of those patterns rather than asking candidates to perform unpaid production work. Customer data, credentials, exploitable vulnerabilities, and proprietary implementation details are removed. Candidate output must not be shipped to production or used to complete an unresolved commercial ticket. The team maps each pattern to the activity, required knowledge, normal tools, expected ownership, consequence of error, and whether the competency must exist at entry or can be learned during onboarding. Repeated activities receive greater weight than exceptional ones; speculative technologies and possible reorganizations are excluded.

The blueprint then defines the minimum evidence for each of the three gates. The first gate lists the smallest set of entry knowledge and experience required for the role. The second selects representative current or upcoming tickets that a candidate should be able to approach. The third selects a product and project artifact through which codebase comprehension and integration readiness can be observed. Topics unsupported by these artifacts are out of scope.

Because backlogs evolve, the blueprint should be regenerated before each hiring campaign or at a declared review interval. It should then be frozen for candidates competing in the same cohort. Updating questions every day while a cohort is active would improve recency at the cost of comparability; versioning preserves evolution across campaigns and consistent treatment within one decision set.

\subsection{Vacancy Artifact}

The framework produces a concise vacancy centered on the position rather than the employer. It identifies principal responsibilities, current stack, representative upcoming work, ownership boundaries, normal tools, minimum entry readiness, and capabilities that may be learned after joining. If a minimum experience range is used, the vacancy states the kind of experience required and does not treat years as an automatic substitute for evidence. It also publishes the number and expected duration of stages, the three decision gates, the early-stop rule, permitted tool and AI use, and any demographic rule that restricts eligibility or can affect selection.

\subsection{Default Assessment Sequence}

TBPAF uses no more than two live sessions by default. The numerical limits below are initial design targets to be evaluated, not empirically established optima.

\textit{Stage 1: minimum-readiness confirmation (20--30 minutes).} A trained assessor checks only the first gate. The candidate describes directly relevant experience and answers a small number of work-grounded questions tied to the declared entry threshold. The assessor asks for a concrete example and one standardized clarification when an answer is uncertain. The session may end when the candidate acknowledges lacking a declared requirement, cannot provide any relevant evidence for a material experience claim after clarification, or otherwise clearly remains below the published minimum. The outcome is communicated without continuing through questions that cannot change the decision.

\textit{Stage 2: ticket and project session (60--90 minutes).} The first part addresses the second gate. The candidate receives one or more sanitized ticket patterns and explains how they would investigate and deliver the work using the normal tool environment. They may demonstrate part of the approach in an editor or live code when useful, but the framework does not award points merely for producing more code under observation. The second part addresses the third gate. The candidate examines a compact product requirement and a related code, legacy, log, test, or architecture artifact; identifies how the pieces connect; asks for missing context; and proposes safe first steps. If all three gates are met, the remaining time is used for mutual discussion of working arrangements, team practices, expectations, and candidate questions.

The default process therefore contains 80--120 minutes of live assessment across at most two appointments. Another stage requires a written explanation of which of the three gates lacks evidence, why the existing session cannot supply it, and the additional time imposed on candidate and staff. Seniority changes the expected ownership, ambiguity, risk, and depth of explanation rather than the number of trivia questions or the duration of the interview.

\subsection{Authentic Tools and Evidence Forms}

The session provides an IDE or equivalent editor, documentation, tests, debugging facilities, search, and representative services used in the job. Internet and AI access follow the advertised working policy. When AI is routinely available, the candidate may use it and is assessed on context construction, evaluation, error detection, security, and adaptation of output. The employer should provide an equivalent sandbox and must not require a personal subscription, private account, or disclosure of a candidate's unrelated prompts and history. A restriction is justified only by an actual unaided responsibility of the position.

Explanation is the default evidence form because it permits candidates to connect experience, ticket reasoning, and project comprehension without turning every role into a timed coding contest. Diagrams, pseudocode, code, tests, or a live demonstration may strengthen or clarify that evidence. They become mandatory only when the blueprint identifies the demonstrated activity itself---rather than knowledge about it---as essential at entry. An optional demonstration should not erase otherwise sufficient evidence because of a minor syntax error or unfamiliar interview environment.

\subsection{Assessor and Computer Responsibilities}

The primary assessor is a developer or manager familiar with the role who has completed basic training and a calibration exercise. The assessor follows the blueprint, uses standardized clarifications, records evidence for each gate, and makes an initial judgment before discussing the candidate with others. A second assessor is added only when independent evidence is necessary, not to create another round.

Computer support presents the same artifact versions, records permitted work products, runs builds or tests when relevant, and connects observations to the three gates. With candidate notice and an appropriate retention policy, it may produce a transcript or structured evidence summary for human verification. It must not score appearance, accent, eye contact, personality, answer smoothness, or similarity to previous successful candidates. It must also avoid treating question count or keyword overlap with model answers as competence. Its purpose is to reduce interviewer-specific variation and preserve evidence, while the accountable hiring decision remains reviewable by a trained person.

\subsection{Implementation as an Assessment System}

TBPAF can be implemented as a company service with five bounded components. A role-analysis component imports the approved vacancy, sanitized backlog patterns, recent completed work, and near-term roadmap; clusters recurring activities; and drafts the versioned blueprint. An artifact builder converts approved patterns into resolved or synthetic tickets, project fragments, and behavioral scoring anchors. A candidate portal publishes the evidence contract: responsibilities, stages, expected time, tools, AI policy, three gates, and any lawful selection rule. A session workspace provides the declared development environment and records only the artifacts and evidence candidates were told would be retained. Finally, an evidence ledger links each assessor observation to one gate, records standardized clarifications and exceptions, and produces an auditable decision packet.

Generation does not equal authorization. A trained hiring owner must approve the blueprint, artifacts, thresholds, and cohort version before use. The system may identify missing evidence, duplicated stages, inconsistent scoring, total candidate time, interviewer interruptions, and group-level outcome patterns; it may not invent requirements, infer protected characteristics, score personality, or issue an unreviewed rejection. This boundary turns automation into process control rather than an opaque substitute for judgment.

\subsection{Three-Gate Decision Rule}

Each gate is rated \textit{not demonstrated}, \textit{minimum demonstrated}, \textit{ready}, or \textit{strong evidence}. Behavioral anchors are generated from the role blueprint and calibrated to seniority. Prior identical employment is one possible source of evidence, not a universal prerequisite: junior candidates, career changers, and candidates from adjacent stacks may demonstrate the required capability through projects, learning evidence, or performance on the supplied artifact. Memorized definitions count only when the candidate can connect them to relevant experience, a ticket, or the supplied project artifact. Interviewer preference for a technology or solution absent from the blueprint has no scoring weight.

All three gates must reach at least \textit{minimum demonstrated}. A failure can trigger early termination only after the standardized clarification is recorded. A material inconsistency between claimed experience and every available concrete example may also stop the process, but the record should describe the unsupported evidence rather than make an unverified judgment about intent. The final outcome is \textit{proceed}, \textit{do not proceed}, or \textit{insufficient evidence}; the last permits one narrow clarification rather than a new general round.

After technical eligibility is established, employment-alignment discussion remains separate from the gate score. Legitimate job constraints such as schedule, location, travel, or on-call availability may inform mutual acceptance when disclosed in the vacancy. Work--life preferences, communication style, motivations, and candidate questions support a two-way decision but should not become an unstructured proxy for cultural similarity. Unrelated private information and protected characteristics are outside the assessment unless a lawful demographic rule was disclosed before application. The final record contains the blueprint and artifact versions, three gate ratings, cited evidence, clarifications, exceptions, and reason communicated to the candidate.

\section{Framework Evaluation and Future Work}
\label{sec:framework_evaluation}

TBPAF is offered as a prescriptive reference framework: its theoretical contribution is the traceable relationship it defines among work, evidence, and decision. Empirical evaluation remains valuable because particular implementations can satisfy the framework on paper yet fail in operation. Evaluation should therefore test both mechanism fidelity and outcomes, examining decision quality and cost together. A shorter process is not an improvement if it produces unreliable or systematically unfair decisions, just as a reliable process is not efficient if it collects the same evidence repeatedly.

\subsection{Replication and Measurement Extension}

An independent replication should preserve the same target construct: remembered professional history compared from one present-day frame. Repeating the instrument in different engineering communities, regions, and future collection periods would show whether the four-era pattern is stable or community-specific. Experience, geography, role family, and seniority can be collected as stratification variables without restricting historical judgment only to firsthand participation. Psychological toll and time commitment should be separated so later studies can identify whether they move together. The era-specific anchors should remain available because they define the contextual comparison; an additional neutral-anchor condition may measure how much detail respondents require to access the same remembered history.

The question benchmark can be extended with item-level source, access date, knowledge domain, project analogue, and independent second coding. Agreement and adjudication would quantify coding stability while preserving the present general-practice question. Vacancy-level matching belongs to implementation of the role blueprint, where the question is whether a generally legitimate topic belongs in one actual selection process; it is not a replacement for RQ4's corpus-level analysis.

\subsection{Comparative Field Study}

A prospective, multi-organization field study should compare TBPAF with each organization's existing process. Requisitions could be randomized where operationally feasible or introduced through a stepped-wedge design in which teams adopt the framework at different times. Comparisons should be stratified by role family and seniority. The unit of assignment should be the vacancy or hiring cohort rather than allowing interviewers to select the method for individual candidates, which would create substantial selection bias.

The primary process outcomes should include candidate preparation time, live-interview time, number of appointments, elapsed days, withdrawal rate, and combined candidate burden measured through separate psychological and time items. Organizational outcomes should include interviewer hours, number and placement of calendar interruptions, scoring time, and time to decision. Candidate-experience measures should cover clarity, perceived job relatedness, tool authenticity, procedural fairness, and quality of feedback.

Decision-quality outcomes require longer follow-up. Inter-rater agreement can be assessed by having independent trained assessors score a common subset of recorded performances. Criterion validity can be estimated against predefined 90- and 180-day work outcomes, such as supervisor ratings anchored to the same role blueprint, ticket completion quality, defect escape, review quality, and team collaboration. No single productivity measure should determine success, and evaluators should account for onboarding, task allocation, and team conditions. False-negative analysis may use later performance of candidates hired elsewhere or structured reconsideration of borderline cases where such data can be obtained ethically.

\subsection{Fairness and Automation Audit}

Stage-by-stage progression, scores, early stops, candidate withdrawals, and offers should be examined across demographic groups where collection is lawful and voluntary. The audit should distinguish access to interview, technical scoring, and any separately disclosed demographic rule. Automated summaries or evidence links should be compared with human-reviewed source material for omission and differential error. Candidates and assessors should be able to flag incorrect records, and performance should be reported both with and without computer assistance.

\subsection{Ablation and Sensitivity Tests}

The framework contains several mechanisms, so a favorable overall result would not show which one caused it. Evaluation should separately vary ticket-derived tasks versus generic tasks, explanation with optional demonstration versus mandatory live coding, authentic versus restricted tools, trained versus uncalibrated assessors, one versus two sessions, and automated evidence support versus manual recording. Sensitivity analysis should test different session limits and minimum thresholds for the three gates. The early-stop rule requires particular scrutiny because time savings must be weighed against false rejection, unsupported inferences about candidate honesty, and demographic differences in how candidates describe uncertainty.

\subsection{Criteria for Adoption}

Adoption should require non-inferior decision reliability and job-performance prediction alongside a meaningful reduction in total candidate and interviewer work. The framework should also improve perceived job relatedness without increasing adverse outcome differences or candidate withdrawal. Results should be reported by role rather than only in aggregate: a process effective for application developers may not transfer to security, infrastructure, research, management, or entry-level positions. These criteria govern organizational adoption and comparison with existing processes; they do not reduce TBPAF to a claim about one pilot implementation. The framework remains a prescriptive, testable design theory, while evidence from particular deployments determines where its mechanisms work, fail, or require adaptation.

\section{Threats to Validity}
\label{sec:threats}

\subsection{Construct Validity}

The survey reconstructs remembered professional history rather than contemporaneously observed employer records. Recency, nostalgia, reputation, and reconstruction consequently shape the construct as well as its uncertainty: the results describe how eras persist in professional memory, not the exact administrative average of every company operating in each year. Live-interview duration and rounds were collected in ranges and represented numerically, so their means depend on midpoint choices and the assigned values for open-ended categories. The burden item combines psychological toll and time commitment and cannot separate them. Era-specific disconnect examples deliberately anchor historically recognizable tensions rather than form a wording-invariant psychometric scale. Every retained respondent selected the same category for all four disconnect dimensions within an era; the aggregate is therefore one contextualized era rating. The 184 distinct cross-era trajectories and distributions in Table~\ref{tab:disconnect_distribution} show that this within-era equality is not uniformity across respondents or history.

The question benchmark measures whether a prompt has a plausible analogue in broad software-engineering practice. It does not measure frequency, importance, predictive validity, appropriateness for a vacancy, or the cumulative composition of a real interview. Coding was performed manually without reported independent raters or inter-rater reliability. Although duplicates were audited, the resulting scores remain an exploratory judgment rather than a validated scale.

\subsection{Internal Validity}

Differences among eras do not by themselves identify industry growth, remote work, AI, applicant volume, demographic policy, or another mechanism in Section~\ref{sec:discussion} as the sole cause. Respondents compared all eras at one point in time, so the pattern combines recollection, professional observation, reputation, nostalgia, recency, and the contextual anchors. Unequal direct exposure is expected for a community-history construct: some judgments come from participation and others from the professional account learned and retained across the field. Repeated structure may encourage comparative consistency, while the non-monotonic disconnect result shows that respondents did not merely increase every measure mechanically.

The study did not link respondents to employers, vacancies, interviewers, applicant volumes, actual schedules, selection outcomes, or later job performance. Consequently, the proposed explanations concerning fragmented calendars, untrained interviewers, question selection, AI assistance, and hidden decision criteria are evidence-informed interpretations rather than tested mediators of the survey results.

\subsection{Conclusion Validity}

The Friedman tests, Kendall's \(W\), and Holm-adjusted adjacent-era signed-rank tests establish structured differences in the retained ratings, but the numerical coding of ordinal ranges does not create true interval measurement. The large sample can produce small \(p\)-values for operationally minor changes, as illustrated by the small Era~III--IV rounds effect. Sensitivity to alternative coding of open-ended categories remains necessary. Developer--manager comparisons are unadjusted descriptive means and may reflect composition rather than role.

The benchmark source pool was not frequency weighted, and item-level provenance was not retained. It therefore cannot estimate how often the analyzed questions are asked. Two duplicate groups with conflicting scores were excluded, but additional semantic duplicates or inconsistent paraphrase coding may remain. The 80.5\% result is conditional on this pool, deduplication method, broad target role, and single coding process.

\subsection{External Validity}

Recruitment through Reddit and LinkedIn produced a voluntary, uncompensated, self-selected community sample. Identity and employment were not independently verified, and the analytical export lacks experience, geography, industry, company size, education, and demographic variables. The sample is dominated by developers rather than hiring managers and cannot be claimed to represent the global software-engineering population. The published artifact supports reproduction from the final 911-record analytical boundary rather than a new audit of the earlier intake and exclusions. Employment law, demographic policy, interview norms, and access to AI differ across jurisdictions; examples from one country or large technology company do not establish prevalence elsewhere. Public preparation sources may overrepresent questions used by prominent employers or candidates seeking highly competitive roles.

\subsection{Validity of the Proposed Framework}

TBPAF is a prescriptive design theory derived from the findings, not an outcome measured by the survey. Its propositions remain conceptually applicable even though implementations will vary. Ticket backlogs may omit tacit responsibilities, overrepresent immediate maintenance, contain confidential information, or change too quickly to define a durable role. Teams with little formal planning may be unable to produce a reliable blueprint, while strategic, research, incident-response, or leadership work may not be expressible as representative tickets. Freezing a blueprint improves candidate comparability but may reduce alignment when business priorities change during recruitment.

Authentic tool access may advantage candidates already familiar with a particular environment, and AI access may differ by language, disability, subscription, or prior opportunity. Automated evidence extraction can reduce interviewer variation yet introduce common systematic bias or privacy risk. Explanation-based evidence may favor verbal fluency, whereas mandatory demonstration may overemphasize performance under observation. Early termination can save time but can also amplify a mistaken gate threshold, candidate anxiety, cultural differences in self-presentation, or an assessor's premature inference that an unsupported claim was intentionally false. These risks define where implementations require human review, versioning, standardized clarification, disclosure, audit, and adaptation as specified in Sections~\ref{sec:framework} and \ref{sec:framework_evaluation}. They are boundary conditions for the theory rather than reasons to return to an unstructured process.

\section{Conclusion}
\label{sec:conclusion}

This study reconstructed the professional history remembered for technical interviews across four eras from 2005 to 2026 and compared that history with an audited corpus of public interview-preparation prompts. The result is a clear account of procedural expansion. Mean live-interview duration increased from 1.34 hours in Era~I to 5.47 hours in Era~IV, mean rounds increased from 2.50 to 4.64, and combined psychological and time burden increased from 2.78 to 8.05. Corrected adjacent-era comparisons confirm every transition except duration from Era~III to IV. Interview--work disconnect followed a different trajectory: it rose from 1.98 to a peak of 3.69 in Era~III before decreasing to 3.14 in Era~IV. Contemporary interviews are therefore remembered as longer and more burdensome even though relevance improved after its Era~III low point.

Developer and hiring-manager responses do not reveal a sharp contemporary disagreement about process scale. Both groups reported the same Era~IV mean number of rounds, while their duration, disconnect, and burden means differed only modestly. These comparisons are descriptive, but they suggest that procedural expansion is visible from both sides of hiring rather than only to candidates.

The question benchmark sharpens rather than weakens the criticism. Of 945 unique prompts with consistent scores, 80.5\% concerned knowledge classified as closely or frequently applicable to general software-engineering practice. Most analyzed questions are not meaningless. The failure occurs when broadly legitimate questions are selected for the wrong vacancy, accumulated beyond the position's scope, asked under artificial conditions, or interpreted inconsistently by untrained assessors. Individual question relevance is not equivalent to validity or proportionality of the complete process.

In response, the paper contributes TBPAF as a prescriptive, testable design theory organized around three gates: minimum role readiness, readiness for the actual ticket pipeline, and readiness to enter the product and codebase. It derives evidence from advertised responsibilities, current and approved upcoming work, and sanitized project artifacts while preserving normal tools and bounding interview time. Its contribution is the explicit and auditable chain from declared work, to required evidence, to a hiring decision. Implementations can now test its mechanisms and outcomes prospectively against existing practice without treating the framework itself as a result of the retrospective survey.

The central practical implication is evidence efficiency: technical hiring should be judged by the job-relevant evidence it collects per unit of candidate and organizational effort. A defensible process need not test everything an engineer might know. It should disclose the job and every decision rule, collect sufficient evidence about the work the employee will actually perform, and stop when the decision is supported. An interview should be a compressed version of the job, not a parallel profession that candidates must learn merely to obtain it.

\appendices
\section{Survey and Reproducibility Materials}
\label{app:materials}

This appendix records the instrument, transformations, and audit rules used for the reported results. The four repeated blocks are identified in the export by their order and correspond to Era~I (2005--2010), Era~II (2011--2017), Era~III (2018--2022), and Era~IV (2023--2026).

\subsection{Survey Instrument}

The form first asked, ``Which describes your role?'' with two response options: \textit{Software Engineer / Candidate} and \textit{Engineering Manager}. Each era block then contained the following common items.

\begin{enumerate}
\item \textit{Live-interview duration:} ``On average, what was the total duration of the live interview process for a single company? Sum up the actual hours spent talking to or coding in front of interviewers, from the initial screen to the final round.'' Options were $<1$ hour, 1--2 hours, 3--4 hours, 5--6 hours, and 7+ hours.

\item \textit{Rounds:} ``On average, how many separate interview rounds (stages/callbacks) did a single company require before extending an offer?'' Options were 1--2 rounds, 3--4 rounds, and 5+ rounds.

\item \textit{Interview expectation versus work reality:} ``Rate the disconnect between the technical skills/paradigms tested during the interview process versus the actual day-to-day work required on the job.'' Four dimension-specific examples followed the stem. Options were 1 (No Disconnect / Perfect Alignment), 2, 3, 4, and 5 (Extreme Disconnect / Completely Irrelevant).

\item \textit{Burden:} ``How would you rate the overall psychological toll and time commitment of the interview process during this era?'' Response options were the integers 1--10.
\end{enumerate}

The four examples attached to the disconnect item changed by era as follows. They are reproduced because the examples form part of the measured construct.

\textit{Era I (2005--2010).}
\begin{itemize}
\item Role and technology alignment: backend versus frontend mismatch, deep infrastructure versus simple scripting, or spoken language versus actual workplace language.
\item Complexity and paradigms: brain-teasers and logic riddles versus standard business logic, or deep theoretical object-oriented programming versus raw procedural code.
\item Knowledge versus tools: rote memory of memory-management trivia versus using standard libraries, documentation, and a local desktop IDE.
\item Teams and operations: boardroom presentation skills versus day-to-day team communication, chaotic project leading, or early Agile adoption.
\end{itemize}

\textit{Era II (2011--2017).}
\begin{itemize}
\item Role and technology alignment: full-stack expectations versus isolated code maintenance, heavy DevOps setups versus feature coding, or mismatched technology stacks.
\item Complexity and paradigms: early LeetCode/HackerRank algorithmic puzzles versus standard web frameworks, CRUD data flows, or rapid feature delivery.
\item Knowledge versus tools: memorizing syntax for browser sandboxes without autocomplete versus searching online documentation and using local environments.
\item Teams and operations: interview ``vibe checks'' and corporate culture fit versus actual cross-functional communication and rigid Scrum/Sprint loops.
\end{itemize}

\textit{Era III (2018--2022).}
\begin{itemize}
\item Role and technology alignment: whole-system design testing versus writing code for a single microservice, or demanding cloud-infrastructure mastery for a product role.
\item Complexity and paradigms: live data-structure puzzles versus cloud orchestration and API glue code, or corporate system-design rounds versus maintaining legacy code.
\item Knowledge versus tools: reciting distributed-systems terminology under live screen monitoring versus reading internal documentation and using modern debuggers.
\item Teams and operations: standardized behavioral-matrix questions, such as the STAR method, versus real cross-functional alignment and autonomous technical leadership.
\end{itemize}

\textit{Era IV (2023--2026).}
\begin{itemize}
\item Role and technology alignment: demanding full-cycle DevSecOps mastery from developers versus highly siloed specialized work, or multi-language hiring requirements versus reality.
\item Complexity and paradigms: LLM-solvable coding puzzles versus complex legacy-codebase maintenance, system integrations, or decoupled modern systems.
\item Knowledge versus tools: absolute bans on AI assistants and invasive screen tracking during tests versus daily use of AI-agent workflows.
\item Teams and operations: isolated live-communication tests versus navigating highly asynchronous, distributed, remote-first alignment or fractional project management.
\end{itemize}

\subsection{Survey Cleaning and Coding}

The form was posted in two waves through software-engineering communities on Reddit and LinkedIn during approximately June and July 2026. Between waves, descriptive instructions were clarified and typographical errors were corrected without changing the analytical scales or response categories. The two waves produced approximately 1,900 raw submissions. Their CSV exports were consolidated, duplicate and incomplete submissions were removed, and records failing a logical-consistency review were excluded. The retained file contains 911 response rows and 30 columns: regenerated row timestamp, role, and seven fields for each of four eras. The original timestamps were not retained because collection time was not an analytical variable. The timestamp sequence in the consolidated CSV was generated during file preparation and must not be interpreted as the survey field period.

Completeness was a retention criterion, so every analyzed row contains every era response and no imputation was performed. Pseudonymous respondent identifiers were assigned, and roles were coded as 1 for developer/candidate and 2 for engineering manager. The retained de-identified table is the declared analytical boundary: all reported denominators, estimates, tests, and respondent-level distributions begin from these 911 records.

Duration categories were represented as $0.5$, $1.5$, $3.5$, $5.5$, and $7.5$ hours in ascending order. Round categories were represented as 2, 4, and 5. For each disconnect response, the leading integer from 1 to 5 was extracted; the four dimension values were averaged within respondent and era. Audit showed that all four values were identical within every retained respondent--era block, so the average equals each constituent response and is treated as one contextualized era rating. This equality was not produced by averaging or imputation. Across time, the retained table contains 184 distinct disconnect trajectories; only 21 respondents used one unchanged rating in all four eras. The burden response was retained as an integer from 1 to 10. These transformations produced four paired observations per measure for each respondent.

Means, medians, sample standard deviations, and normal-approximation 95\% confidence intervals for the mean were calculated for each era. Each confidence interval used the mean plus or minus 1.96 times the sample standard error. Friedman tests used within-respondent average ranks and a correction for ties. The resulting omnibus statistics were $\chi^2_F(3)=2142.37$ for duration, $1830.44$ for rounds, $1263.72$ for disconnect, and $2259.79$ for burden; corresponding Kendall's \(W\) values were \(0.784\), \(0.670\), \(0.462\), and \(0.827\). Adjacent eras were compared using two-sided Wilcoxon signed-rank tests with zero differences removed, average ranks for ties, normal approximation with tie and continuity corrections, and Holm adjustment across the three comparisons for each measure. Sensitivity checks assigned the 7+ hour category values from 7.0 to 10.0 and the 5+ rounds category values from 6 to 8. Era~IV developer--manager comparisons used unadjusted group means only.

\subsection{Interview-Question Benchmark Audit}

The raw benchmark file contained 1,001 rows with three fields: interview question, closest real-work analogue, and Difference score. These rows were not treated as 1,001 distinct questions. Prompts were stripped of surrounding whitespace and grouped by case-folded exact text, producing 947 unique prompt groups. Repeated prompts with one consistent score contributed one record. Two groups containing more than one score were excluded rather than adjudicated, leaving 945 unique consistently scored prompts for analysis.

The final counts were 410 prompts at score 1, 351 at score 2, 111 at score 3, 46 at score 4, and 27 at score 5. The source pool represents openly accessible interview-preparation material, including LeetCode, HackerRank, public GitHub repositories, and public Reddit discussions; prompts were not collected directly from identifiable individuals. The pool is not frequency weighted. Item-level source URLs were not retained, so the popularity and exact original provenance of individual prompts cannot be reconstructed from the artifact. Public accessibility does not establish public-domain status, and redistribution remains subject to applicable source terms.

\subsection{Reproduction Procedure and Data Availability}

The artifact package contains the retained consolidated survey CSV, a de-identified analytical table, the raw 1,001-record question CSV from which 945 unique prompts were analyzed, and a reproduction script in the repository's scripts directory. Running the script from the repository root reproduces sample counts, era descriptives, tie-corrected Friedman statistics, Kendall's \(W\), adjacent-era Wilcoxon tests, disconnect category distributions and trajectories, Era~IV role means, and the audited benchmark distribution using only the Python standard library.

The consolidated survey file contains regenerated row timestamps rather than original collection times and contains no respondent names. Those timestamps are transport metadata and are excluded from every analysis. A public replication package should distribute the de-identified analytical table by default. The question dataset may be redistributed only to the extent permitted by the terms of its underlying sources.

\bibliographystyle{IEEEtran}
\bibliography{references}

\end{document}